\documentstyle[epsfig,12pt]{article}
                                                                                       
\newlength{\dinwidth}                                                                                       
\newlength{\dinmargin}                                                                                       
\def\lapproxeq{\lower .7ex\hbox{$\;\stackrel{\textstyle                                                                                       
<}{\sim}\;$}}                                                                                       
\def\gapproxeq{\lower .7ex\hbox{$\;\stackrel{\textstyle                                                                                       
>}{\sim}\;$}}

\def\msb{\overline{\rm MS}}

\def\beq{\begin{equation}}                                                    
\def\eeq{\end{equation}}                                                    
\def\bea{\begin{eqnarray}}                                                    
\def\eea{\end{eqnarray}}                                                   
\def\be{\begin{equation}}                                                                                       
\def\ee{\end{equation}}                                                                                       
\def\bea{\begin{eqnarray}}                                                                                       
\def\eea{\end{eqnarray}}                                                                                       
\def\sml{\textstyle}
                                                                                       
\begin{document}                                                                                       
\titlepage                                                                                       
\begin{flushright}                                                                                       
RAL-TR-1999-028 \\  
April 1999                                                                                       
\end{flushright}                                                                                       
                                                                                       
\begin{center} 
\vspace*{2cm}                                                                                       
{\Large \bf Analysis of deep inelastic scattering with \\ [2mm]
$z$-dependent scale } \\
\vspace*{1cm}            
R.\ G.\ Roberts\\                                                                                       
                                                                                       
{\it Rutherford Appleton Laboratory, Chilton,                                                                                       
Didcot, Oxon, OX11 0QX}\\                                                                                       
\end{center}                                                                                       
                                                                                       
\vspace*{1.5cm}                                                                                       
\begin{abstract}  
Evolution of the parton densities at NLO in $\alpha_S$ 
 using $\tilde W^2 = Q^2 (1-z)/z$ instead
of the usual $Q^2$ for the scale of the running coupling $\alpha_S$ is
investigated. While this renormalisation scale change was originally proposed
as the relevant one for $x \rightarrow 1$, we explore the consequences for
all $x$ with this choice. While it leads to no improvement to the description 
of DIS data, the nature of the gluon at low $x$, low $Q^2$ is different,
avoiding the need for a `valence-like' gluon. 
\end{abstract}                                                

\newpage

\noindent{\large \bf 1. Motivation $-$ Non-singlet evolution}

In deep inelastic scattering (DIS) the $Q^2$ dependence of flavour
non-singlet quantities is quite straightforward.
Taking a non-singlet structure function, e.g.
$ F_{NS}^+ = F_2^{ep} - F_2^{en}$ or 
$F_{NS}^- =F_2^{\bar \nu p} - F_2^{\nu p}$
then at next-to-leading order (NLO) we have
\beq
F_{NS}^{\pm}(x,t) = \tilde q_{NS}^{\pm}(x,t) + 
\left ( \frac{\alpha_S}{4\pi} \right )B_q (z) \otimes 
\tilde q_{NS}^{\pm}(x/z,t) 
\label{eq:NS1}
\eeq
where $t = \ln Q^2 $ , $ \tilde q(x,t) = x q(x,t)$, the relevant
combination of quark and antiquark densities (weighted by the appropriate
charges squared)
 is denoted by $q_{NS}$
and $B_q(z)$ is the quark coefficient function given, for example, in
 the $\msb$ scheme by 
\beq
B_q(z) = \left [ \hat B_q(z) \right ]_+ =
\left [ \hat P_{qq}^{(0)}(z) \left \{ \ln \left ( \frac{1-z}{z} \right )
- \frac{3}{2} \right \} + {1\over 2} (9+5z) \right ] _+
\label{eq:Bq}
\eeq
where 
\beq
P_{qq}^{(0)}(z) = [\hat P_{qq}^{(0)}(z)]_+ = 
 \left [ C_F \left ( \frac{1+z^2}{1-z} \right ) \right]_+
\label{eq:PNS}  
\eeq
is the well-known LO $q$-$q$ splitting function.

Here $[...]_+$ denotes the standard regularised functions defined by
\beq
\int_0^1 dz \; [f(z)]_+ \; g(z) = \int_0^1 dz \; f(z)\;[g(z)-g(1)]
\label{eq:plus}
\eeq

The NLO evolution of the non-singlet quark density is governed by the 
$qq$ and $\bar q q$ splitting functions
\bea
\frac{d }{dt} \tilde q_{NS}^{\pm}(x,t) &=& 
\left \{ \left [ \left ( \frac{\alpha_S}{2\pi} \right )
\hat P_{qq}^{(0)}(z) +  
\left ( \frac{\alpha_S}{2\pi} \right )^2
 \hat P_{qq}^{(1)}(z)   
 \right ]_+  \pm \left ( \frac{\alpha_S}{2\pi} \right )^2
 \hat P_{\bar qq}^{(1)}(z) \right . \nonumber \\
&+& \left . \delta (1-z) \int_0^1 dz \left ( \frac{\alpha_S}{2\pi} \right )^2
\hat P_{\bar qq}^{(1)}(z) \right \}
\: \otimes \: \tilde q_{NS}^{\pm}(x/z,t) 
\label{eq:apns}
\eea

The NLO splitting functions take the form
\bea
\hat P_{qq}^{(1)}(z) &=& C_F^2\;P_F(z) + \frac{1}{2}C_F C_A P_G(z)
+ \frac{1}{2} C_F N_F P_{N_F}(z) \nonumber \\
\hat P_{\bar qq}^{(1)}(z ) &=& (C_F^2 - \frac{1}{2} C_F C_A) P_A(z)
\label{pqqnlo}
\eea
and the explicit expressions for $P_F(z)$, $P_G(z)$, $P_{N_F}(z)$ and
$P_A(z)$ can be found in \cite{cfp} for example.

By combining eqs.(\ref{eq:NS1},\ref{eq:apns}), 
the evolution of the non-singlet structure functions to ${\cal O} 
(\alpha_S^2)$ may then be expressed in the form
\bea
\frac{d }{dt} F_{NS}^{\pm}(x,t) &=& 
\left \{ \left [ \left ( \frac{\alpha_S}{2\pi} \right )
\hat P_{qq}^{(0)}(z) +  
\left ( \frac{\alpha_S}{2\pi} \right )^2
\large \{ \hat P_{qq}^{(1)}(z)   
- {\sml \frac{1}{4}} \beta_0 \hat B_q(z) \large \}
\right ]_+ \pm \left ( \frac{\alpha_S}{2\pi} \right )^2
 \hat P_{\bar qq}^{(1)}(z)\right . \nonumber \\
&+& \left . \delta (1-z) \int_0^1 dz \left ( \frac{\alpha_S}{2\pi} \right )^2
\hat P_{\bar qq}^{(1)}(z) \right \}
\: \otimes \: F_{NS}^{\pm}(x/z,t) 
\label{eq:apns2}
\eea

If, in the above $\alpha_S$ is a function of $t$ only, the running
coupling can be taken outside the convolution integral and 
we have
the usual NLO in $\alpha_S$ evolution as implemented in standard 
analyses of DIS. 

Next consider the case where the relevant scale depends on
$z$ as well  i.e. $\alpha_S(Q^2) \rightarrow \alpha_S(\phi(z)Q^2)$
and, in particular, the choice $\phi(z)=(1-z)/z$. 
The quantity $\tilde W^2 = Q^2(1-z)/z$ is just the com energy squared  
of the virtual photon-parton
scattering which controls the maximum tranverse momentum occuring in
the ladder graphs which are summed to give the leading log contribution.
As $z \rightarrow 1$, $\tilde W^2$ as well as $Q^2$ is large and it has 
been argued \cite{bl,abcm} that, in this region, the relevant 
variable to account for the large logs which arise beyond the control
of the renormalisation group is $W^2$ or a quantity closely related.  

Expanding $\alpha_S(t + \ln [\phi(z)] )$ to ${\cal O} (\alpha_S^2)$ we get
\beq
\frac{\alpha_S(t + \ln \phi(z) )}{2\pi} = \frac{\alpha_S(t)}{2\pi} - 
{\sml \frac{1}{2}} \beta_0 \ln [\phi(z)] 
\left ( \frac{\alpha_S(t)}{2\pi} \right )^2  
\label{eq:alphaexp}
\eeq
which means that, to this order, the change of scale
of $\alpha_S$ is equivalent to adding an extra term inside
the $[...]_+$ of eq.(\ref{eq:apns}) equal to
\beq
{- \sml \frac{1}{2}} \beta_0 \left ( \frac{\alpha_S(t)}{2\pi} \right )^2
\hat P_{qq}^{(0)}(z) \ln [\phi(z)]
\label{eq:extraterm}
\eeq

The above expressions for the evolution for $\tilde q_{NS}^{\pm}$ and
$F_{NS}^{\pm}$ are now to be understood with $\alpha_S \equiv \alpha_S(t,z)$
but we can see from eqs.(\ref{eq:apns2},\ref{eq:extraterm}) that
to ${\cal O} (\alpha_S^2(t))$ the shift in scale is equivalent to a term
in $\hat B_q(z)$ equal to $2 \hat P_{qq}^{(0)}(z) \ln [\phi(z)]$ or a
term in $\hat P_{qq}^{(1)}(z)$ equal to 
$-{\sml \frac{1}{2}} \beta_0 \hat P_{qq}^{(0)}(z) \ln [\phi(z)]$. Thus taking
$\phi(z)=(1-z)/z$ the first term in eq.(\ref{eq:Bq}) is generated by the
change of scale $Q^2 \rightarrow Q^2(1-z)/z$ and then to avoid double
counting at ${\cal O}(\alpha_S^2(t))$ we should use the simpler form 
\beq
\hat B_q(z) \longrightarrow \;-{\sml \frac{3}{2}}\hat P_{qq}^{(0)}(z)    
+{\sml \frac{1}{2}}(9+5z)
\label{eq:Bq2}
\eeq

Since $\beta_0 = {\sml\frac{1}{4}}C_F(11C_A-2N_F)$ we note that
some of the $N_F$
dependence in the combination of the NLO NS splitting function and  the
quark coefficient function has been absorbed by the change of scale. 
One can go further and compute the form $\phi(z)$ should take in order to
generate the entire $N_F$ dependence at ${\cal O} (\alpha_S^2(t))$.
Wong\cite{wong} showed that can be achieved by taking
\beq
\phi(z) \rightarrow \tilde \phi(z) = \left ( \frac{1-z}{z^2} \right )
\exp  [ (C_F/\hat P_{qq}^{(0)}(z))({\sml \frac{1+13z}{4}}) -
{\sml \frac{29}{12}} ]
\label{eq:wongphi}
\eeq  

This is the BLM procedure\cite{blm} where the $N_F$ dependence 
in a particular process is 
identified as arising from the vacuum polarisation contributions
to that process and these are then entirely absorbed into the running
coupling thus providing a method of summing such contributions to all orders.
This procedure is therefore quite attractive but it requires the choice
of the $z$-dependent scale to vary from process to process. In that sense
the evolution of the non-singlet structure function is a different process
from the singlet one. It is impractical to attempt an analysis of DIS
where the choice of renormalisation scale differs for the singlet and 
non-singlet combinations and here we shall explore the 
the consequences for a single choice of scale, $\phi(z)=(1-z)/z$.
It is clear that this simple choice accounts, at ${\cal O} (\alpha_S^2)$,
for large terms in the $\msb$ coefficient functions ocurring both in
singlet and non-singlet expressions and it is worth investigating the
phenomenological consequences of the potentially large logarithms
which are summed by this procedure over the entire range of $x$.

\noindent{\large \bf 2. Quark singlet and gluon evolution}

  Consider now the evolution of a singlet structure function, 
$F_S(x,t)$
where 
\beq
F_{S}(x,t) = \tilde q_{S}(x,t) + 
\left ( \frac{\alpha_S}{4\pi} \right )B_q (z) \otimes 
\tilde q_{S}(x/z,t) + 
\left ( \frac{\alpha_S}{4\pi} \right )B_g (z) \otimes 
\tilde g(x/z,t) 
\label{eq:S1}
\eeq
where $ \tilde q_S(x,t) = x \sum (q(x,t)+\bar q(x,t))$, 
and $B_g(z)$ is the gluon coefficient function given in
 the $\msb$ scheme by 
\beq
B_g(z) =  2 \hat P_{Sg}^{(0)}(z) \ln \left ( \frac{1-z}{z} \right )
+ 2 N_F [8z(1-z) -1]
\label{eq:Bg}
\eeq
where 
\beq
\hat P_{Sg}^{(0)}(z) = N_F\;[z^2 + (1-z)^2\;]
\label{eq:PSG}  
\eeq
is the LO $q$-$g$ splitting function.

The NLO evolution of the singlet quark density is governed by the 
appropriate $SS$ and $Sg$ splitting functions
\bea
\frac{d }{dt} \tilde q_{S}(x,t) &=& 
\left \{ \left [ \left ( \frac{\alpha_S}{2\pi} \right )
\hat P_{qq}^{(0)}(z) +  
\left ( \frac{\alpha_S}{2\pi} \right )^2
 \hat P_{qq}^{(1)}(z)   
 \right ]_+  + \left ( \frac{\alpha_S}{2\pi} \right )^2
\left \{ \hat P_{\bar qq}^{(1)}(z) + \Delta \hat P_{qq}^{(1)} (z) \right \}
\right . \nonumber \\
&+& \left . \delta (1-z) \int_0^1 dz \left ( \frac{\alpha_S}{2\pi} \right )^2
\hat P_{\bar qq}^{(1)}(z) \right \}
\: \otimes \: \tilde q_{S}(x/z,t) \nonumber \\
 &+&\left \{  \left ( \frac{\alpha_S}{2\pi} \right )
\hat P_{Sg}^{(0)}(z) +  
\left ( \frac{\alpha_S}{2\pi} \right )^2
 \hat P_{Sg}^{(1)}(z)   
 \right \}  \: \otimes \: \tilde g(x/z,t)  
\label{eq:aps}
\eea
where $\Delta \hat P_{qq}^{(1)}(z)$ and $\hat P_{SG}^{(1)}(z)$ have the form
\bea
\Delta \hat P_{qq}^{(1)}(z) &=& C_F\;N_F\;F_{qq}(z) \nonumber \\
\hat P_{SG}^{(1)}(z) &=& {\sml \frac{1}{2}}C_A N_F F_{qg}^{(1)}(z) +
{\sml \frac{1}{2}}C_F N_F F_{qg}^{(2)}(z)
\label{eq:Ffuns}
\eea
and the relevant expressions for the $F$'s can be read off from 
ref.\cite{fp}. 
As in the non-singlet case, we can combine eqs.(\ref{eq:S1},\ref{eq:aps})
and the evolution of the gluon into a `one-step' evolution of the 
singlet structure function\cite{hww} correct to ${\cal O}(\alpha_S^2)$
which is of the form
\beq
\frac{d}{dt}
\left ( \begin{array}{c} F_S(x,t) \\ \tilde g(x,t) \end{array} \right ) =
\left \{ \left ( \frac{\alpha_S}{2\pi} \right )
 {\cal P}_S^{(0)}(z) +  
\left ( \frac{\alpha_S}{2\pi} \right )^2
\left \{ {\cal P}_S^{(1)}(z)-\frac{1}{4} \beta_0 {\cal D}^{(1)}(z)
+{\cal E}^{(1)}(z) \right \} \right \} \otimes
\left ( \begin{array}{c} F_S(x/z,t) \\ \tilde g(x/z,t) \end{array} \right )
\label{eq:onestep}    
\eeq
where ${\cal P}_S^{(0)}$, ${\cal P}_S^{(1)}$, ${\cal D}^{(1)}$ and
${\cal E}^{(1)}$ are given in ref~\cite{hww}.

Using eq.(\ref{eq:alphaexp}) we have
again that the effect of changing the scale $Q^2 \rightarrow Q^2(1-z)/z$
is to generate, at ${\cal O}(\alpha_S^2)$, the logarithm terms which 
explicitly appear in both the $\msb$ gluon and quark coefficient
functions. So in addition to eq.(\ref{eq:Bq2}) we take
\beq
B_g(z) \longrightarrow  2 N_F [8z(1-z) -1]
\label{eq:Bg2}
\eeq

For the gluon evolution we simply insert counter
terms involving $\phi(z)$ to restore eq.(\ref{eq:onestep})
at {\cal O}$(\alpha_S^2(t))$ generated by the scale change.
\bea
\frac{d}{dt} \tilde g(x,t) &=&
\left \{ \frac{1}{z} \left [ \left ( \frac{\alpha_S}{2\pi} \right )
\; z  \hat P_{gg}^{(0)}(z) + \left ( \frac{\alpha_S}{2\pi} \right )^2
\; z  \left ( \hat P_{gg}^{(1)}(z) + {\sml \frac{1}{2}} \; \beta_0 \;
\hat P_{gg}^{(0)}(z) \ln [\phi(z)] \right )
\right ]_+ \right . \nonumber \\
 &-& \left . \delta(1-z) \int_0^1 dz 
\left [ \left ( \frac{\alpha_S}{2\pi} \right )
\hat P_{Sg}^{(0)}(z) + \left ( \frac{\alpha_S}{2\pi} \right )^2
\hat P_{Sg}^{(1)}(z) \right ] \right \} 
\otimes \tilde g(x/z,t) \nonumber \\
 &+& \left \{ \left ( \frac{\alpha_S}{2\pi} \right ) \hat P_{gS}^{(0)}(z)
+ \left ( \frac{\alpha_S}{2\pi} \right )^2 \left (\hat P_{gS}^{(1)}(z)
+ {\sml \frac{1}{2}} \; \beta_0 \;
\hat P_{gS}^{(0)}(z) \ln [\phi(z)]  \right ) \right \} \otimes \tilde q_S(x/z,t)
\label{eq:apg}
\eea

While in the non-singlet case a single change of scale for $\alpha_S$ 
(eq.(\ref{eq:wongphi})) can be chosen to absorb all the $N_f$ dependence
at ${\cal O}(\alpha_S^2)$, the situation is more complicated here and a
similar BLM procedure would involve different choices of scale for the 
quark singlet and gluon evolution. An alternative to eq.(\ref{eq:onestep})
would be a one step evolution expressed in terms of physical structure
functions only, $F_S(x,t)$ and $F_L(x,t)$ in which case the modifications
due to $Q^2 \rightarrow \phi (z) Q^2$ could be entirely absorbed at 
${\cal O}(\alpha_S^2)$ into the coefficient functions $B_q(z)$ and $B_L(z)$. 

\noindent{\large \bf 3. Fitting the DIS data}

A practical problem that arises in using $\phi(z)=(1-z)/z$ is simply that
the integrations involved in the convolution integrals run up to $z=1$
and so we must adopt some sensible approach for computing the running
coupling at low values of the scale. Since the argument $\tilde W^2$
of $\alpha_S$ is actually timelike, then beyond leading order
the coupling is complex. (Recall that the perturbation expansion
is derived strictly in spacelike region and then continued analytically
to the timelike region.)  
We consider two possibilities (i)
compute $\alpha_S(|\tilde W^2|)$ in terms of $\Lambda_{\msb}$ in the standard
NLO way but `freeze' its value for $|\tilde W^2| < Q^2_f$ 
and (ii) use $|\alpha_s(\tilde W^2)|$ as the running coupling. 
This latter variable
is claimed\cite{pr} to be a far more efficient expansion parameter 
(since it resums large terms involving $\pi^2$ arising from the
analytic continuation) and
grows only weakly as the argument becomes very small. The imaginary
part of $\alpha_S$ is computed from the modulus at NLO,
\be
{\rm Im} (\alpha_S) = \frac{\beta_0}{4} |\alpha_S|^2
+ \frac{\beta_1}{16\pi} |\alpha_S|^3   
\label{eq:ima}
\ee

\begin{figure}[H] 
\vspace{-3cm}
\begin{center}      
\epsfig{figure=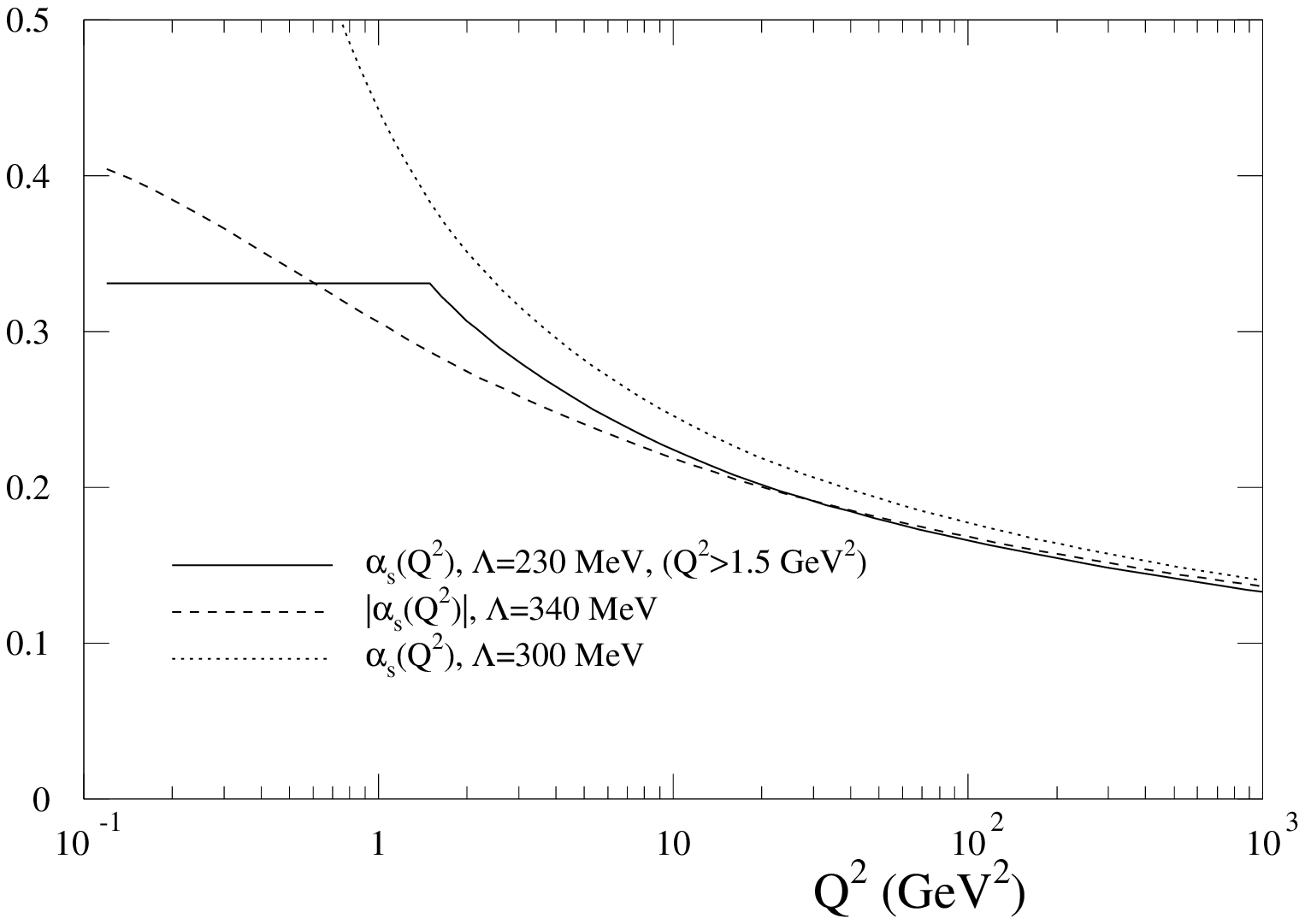,height=20cm}
\end{center}
\vspace{-7cm}     
\caption{Running couplings used in the various fits. The solid line
shows $\alpha_S(|\tilde W^2|)$ versus $\tilde W^2$, the dashed line
shows $|\alpha_S(\tilde W^2)|$ versus $\tilde W^2$, the values of
$\Lambda_{\msb}$ being indicated.  The dotted line shows, for the sake
of comparison versus $\alpha_S(Q^2)$ versus $Q^2$, the running coupling
used in the standard MRST fit.}
\label{fig:fig1} 
\end{figure}

 Fig.~\ref{fig:fig1} displays the running couplings used in our fits
compared with the $\alpha_S$ used by a conventional fit using a 
$z$-independent scale, the values of $\Lambda_{\msb}$ giving the best
fit to the DIS data being shown. For the choice $\alpha_S(|\tilde W^2|)$
the quality of the fit is not overly sensitive to 
the precise value of the scale at which the coupling is frozen and we take
a value around 1 GeV$^2$. 

\begin{figure}[H] 
\vspace{-3cm}
\begin{center}      
\epsfig{figure=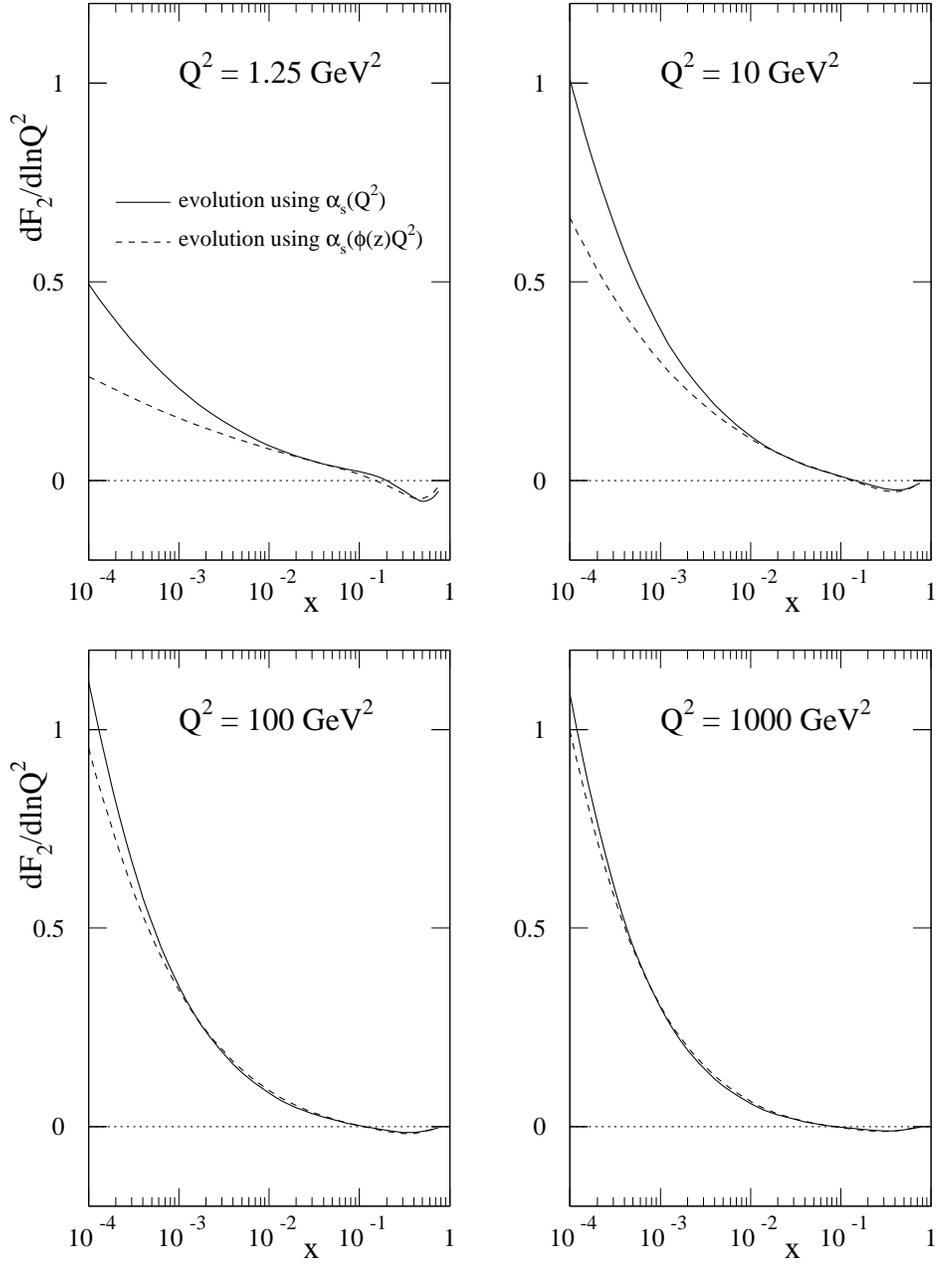,height=20cm}
\end{center}
\caption{Comparison of the evolutions with scales $\tilde W^2$ and $Q^2$
from a common set of parton distributions at $Q^2_0$ = 0.5 GeV$^2$ and
with a common value of $\Lambda_{\msb}$.}
\label{fig:fig2} 
\end{figure}

As with the usual analysis of DIS structure functions, the parton
distribution functions (pdf's) are parametrised at some starting scale, 
$Q^2_0$, but now evolved according to 
eqs.(\ref{eq:apns2},\ref{eq:aps},\ref{eq:apg}). In the
recent MRST fits\cite{mrst98}, the starting scale is $Q^2_0$ = 1 GeV$^2$
and the gluon at small $x$ is suppressed (so-called `valence-like' gluon)
at this $Q^2$ in order to describe the relatively small slope 
$dF_2/d\ln Q^2$ of the structure function observed at HERA\cite{H1,ZEUS}
at low $x$ and low $Q^2$. At low $x$, the mean value of $z$ in the 
evolution is also low implying that the scale $\tilde W^2$ is
much larger than $Q^2$ and hence the gluon evolution is naturally
suppressed at low $x$. In Fig.~\ref{fig:fig2} we show a comparison of
the structure function evolution for a common set of pdf's at $Q^2_0$ 
according to the two scales. The starting scale is $Q^2_0 = 0.5$ GeV$^2$
and it is clear that evolving with the scale $\tilde W^2$ leads to
dramatically slower evolution at low $x$ and low $Q^2$, even with a
common value of $\Lambda_{\msb}$. 

The concept of a gluon distribution vanishing as $x \rightarrow 0$ 
(even at low $Q^2$) seems
unnatural and leads to problems when attempting to evolve down in $Q^2$
since physical quantities, such as $F_L$ quickly become negative. 
The comparison in Fig.~\ref{fig:fig2} shows we can instead start with
a larger gluon and in fact we find the data can be fitted
a gluon distribution which is actually singular for $Q^2_0$ as low as 
0.5 GeV$^2$.

\begin{figure}[H] 
\vspace{-3cm}
\begin{center}      
\epsfig{figure=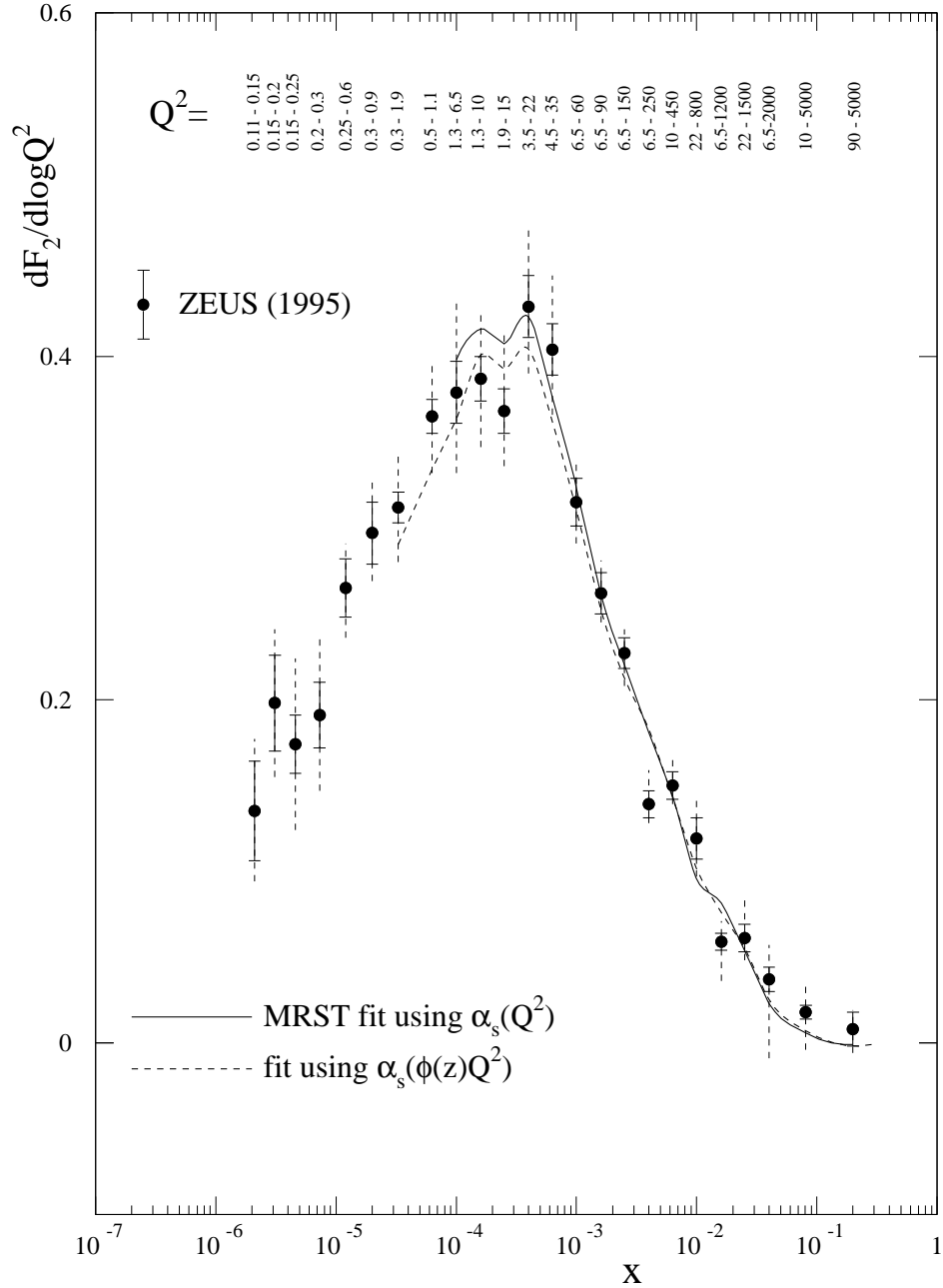,height=20cm}
\end{center}
\caption{Slope of $F_2$ for different $x$ values measured by 
ZEUS\protect\cite{lowzeus}.
Note the strong correlation between the value of $x$ and the mean value of 
$Q^2$. The data are compared to fits to DIS data using a scale $Q^2$
or a scale $\tilde W^2$.}
\label{fig:fig3} 
\end{figure}

DIS data fitted include the HERA data\cite{H1,ZEUS,lowzeus},BCDMS\cite{BCDMS},
E665\cite{E665}, SLAC\cite{SLAC} and NMC\cite{NMC}. We label the two types
of fits as type (i) or type (ii) depending on the choice 
$\alpha_S(|\tilde W^2|)$ or $|\alpha_S(\tilde W^2)|$.
There has been much 
interest in the values of the slope $dF_2/d\ln Q^2$ shown by the low $x$, 
low $Q^2$ 1995 ZEUS data\cite{lowzeus} that has prompted speculation about a
possible breakdown of the standard theory\cite{mueller}. 
A type (i) fit gives a satisfactory
description of the slopes observed by ZEUS without the need to invoke a
`valence-like' gluon distribution at low $Q^2$ $-$ as shown in 
Fig.~\ref{fig:fig3} and as anticipated above. Because of the more conventional
behaviour of the gluon at small $x$ one can evolve to values of $Q^2$
below 1 GeV$^2$. At the starting scale $Q^2_0 = 0.5$ GeV$^2$ the gluon
has the form
\be
xg(x,Q^2_0)= 1.25 x^{-0.046}(1-x)^{5.52}(1+0.032\sqrt{x}+5.66x)
\label{eq:gluon0}   
\ee
which is virtually `flat' at small $x$.

Fig.~\ref{fig:fig3} clearly shows a good description of the $F_2$ slopes
in the HERA range both for the conventional type of fit and for the new
type (i) fit using a $z$-dependent scale. Copmparing in detail the quality
of the fits, we obtain
\begin{center}
\begin{tabular}{|c|c|c|c|c|c|c|}\hline
 & H1 & ZEUS & BCDMS & NMC & SLAC & E665 \\ \hline
No. data pts & 221 & 216 & 174 & 130 & 70 & 53 \\ \hline
$\chi^2$ (MRST) & 164 & 264 & 249 & 143 & 116 & 56 \\ \hline
$\chi^2$ (type (i) fit) & 164 & 270 & 243 & 170 & 134 & 52 \\ \hline
$\chi^2$ (type (ii) fit) & 178 & 298 & 260 & 223 & 156 & 48 \\ \hline
\end{tabular}
\end{center}

\begin{figure}[H] 
\vspace{-3cm}
\begin{center}      
\epsfig{figure=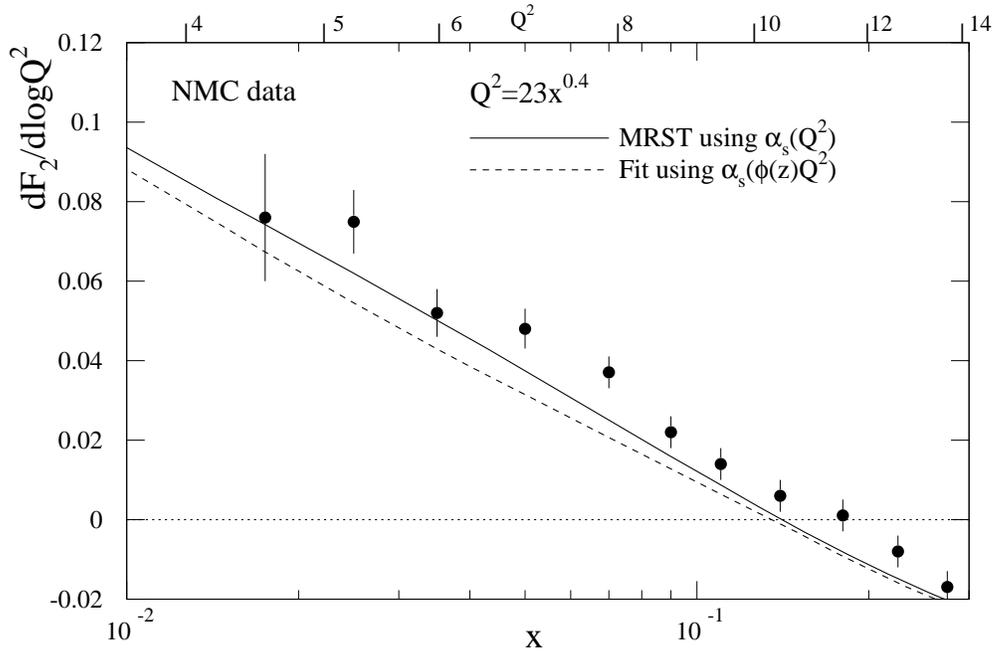,height=20cm}
\end{center}
\vspace{-7cm}     
\caption{Slope of $F_2$ for different $x$ values measured by 
NMC\protect\cite{NMC}.
Again the values of $x$ and the mean value of $Q^2$ are correlated. 
The data are compared to fits to DIS data using a scale $Q^2$
or a scale $\tilde W^2$.}
\label{fig:fig4} 
\end{figure}

In the new fits, the cuts applied are the same as for MRST, in particular
$Q^2 > 2 $ GeV$^2$ except for the HERA data where we take $Q^2 > 1.5$ 
GeV$^2$. 
For the HERA data with $Q^2 > 1.5$ GeV$^2$, fit (i) achieves virtually the
same quality as the conventional type fit but can, in addition, give a good
description down to $Q^2 = 1$ GeV$^2$. The type (ii) is slightly worse for
the HERA data.  For the `intermediate' $x$ region covered by NMC, the new
fits are not as successful as the conventional MRST fit, especially the type
(ii) fit, and this is due to problem of trying to describe the $Q^2$ slopes 
of the NMC data while simultaneously being consistent with the slopes
measured at HERA. The MRST fit already underestimates the slopes 
observed by NMC and the new fits only serve to emphasise the disagreement  
as shown in Fig.~\ref{fig:fig4}. Type (ii) fit being even worse indicates 
that the results are sensitive to the precise nature of the coupling at
very low values of the scale.

\noindent{\large \bf 4. Conclusions}

Although the motivation for evolving partons with a scale $\phi(z) Q^2$
where $\phi(z)=(1-z)/z$ stems from a procedure for resumming terms which
are potentially large as $x \rightarrow 1$, it is worth exploring the
phenomenological consequences for quarks and gluons over the whole $x$
region. In fact we did examine whether the large $x$ region was better
described in terms of a running coupling with the choice of scale
$Q^2(1-z)/z$ - the hope being that the observed strong $Q^2$ observed
at values of $x$ beyond 0.6 might be absorbed by such a choice instead of 
conventional higher twist contribution ( see ref.\cite{MRST2} for example)
but we found no evidence for this. 
The change of scale $Q^2 \rightarrow \phi(z) Q^2$ generates
corrections at ${\cal O}(\alpha_S^2)$ which already appear explicitly
as the logarithm terms in the $\msb$ coefficient functions, so evolving  
with the $z$-dependent scale implies somewhat simpler expressions
for the coefficient functions. It should be remembered that our change
of scale is a change of renormalisation scheme and in order to maintain the
expressions for the evolution of the physical $F_2$ structure function to
${\cal O}(\alpha_s^2)$ we must modify the coefficient functions and the
relevant splitting functions as shown in sections 1 and 2. Also the 
regularising counter terms (proportional to $\delta(1-z))$ are more
complicated due to $\alpha_S$ depending on $z$.  

Performing fits with two alternative prescriptions for handling 
the running coupling at low values of the scale leads to a preference
for the simpler choice of freezing its value at around 1 GeV$^2$. In this
case one can get acceptable fits to the DIS data but no improvement over
the standard procedure is observed; in fact the problem of trying to reconcile
the slopes $dF_2/d\ln Q^2$ measured by NMC with those measured at HERA is
is actually aggravated. This subtle $Q^2$ dependence of the slopes
has only been successfully described when effects beyond standard DGLAP
physics are included. In particular Thorne\cite{rob1} was able to achieve 
a high degree
of consistency between all the DIS data sets with a leading order,
renormalisation scheme consistent calculation which included leading
$\ln (1/x)$ terms. 

\begin{figure}[H] 
\vspace{-3cm}
\begin{center}      
\epsfig{figure=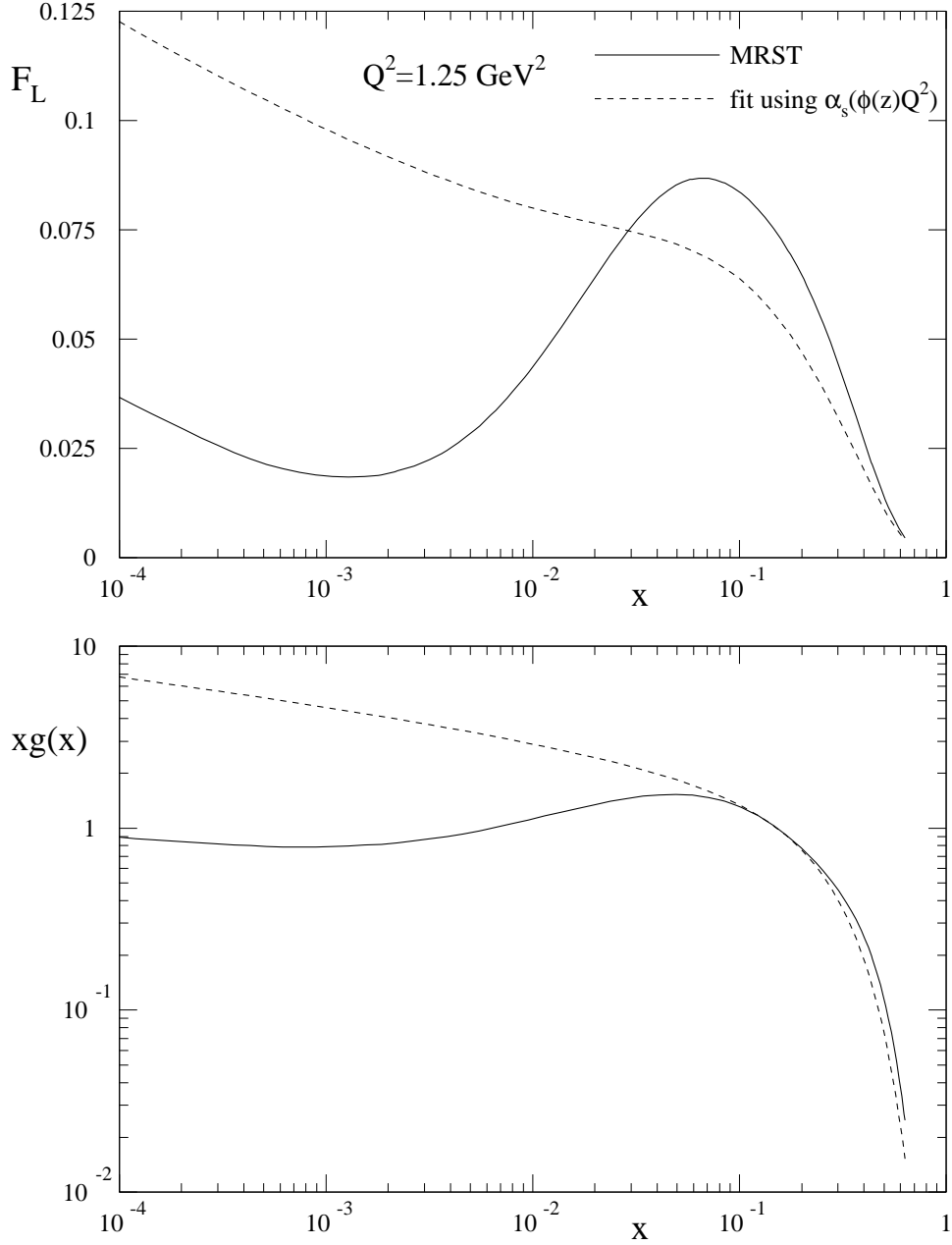,height=20cm}
\end{center}
\caption{A comparison of $F_L(x,Q^2=1.25)$ predicted by two fits, one
(MRST) using the standard evolution scale, the other the type (i) fit
using a scale $\tilde W^2$. Also shown are the respective gluon distributions
at the same $Q^2 = 1.25$ GeV$^2$ which are responsible for driving $F_L$.} 
\label{fig:fig5} 
\end{figure}

An advantage which has been gained in using the scale $Q^2(1-z)/z$ is that
the gluon distribution at low $Q^2$ seems more `natural' than the 
`valence-like' gluon which the standard DGLAP description finds necessary
to account for the small values of the slopes observed by ZEUS\cite{lowzeus}.
We therefore expect that quantities dominantly governed by the gluon
are sensitive to the change of scale, in particular the longitudinal
structure function at low $x$, low $Q^2$ is quite different. 
Fig.~\ref{fig:fig5} shows a comparison between predictions for $F_L$ at
$Q^2=1$ GeV$^2$  
resulting from fits to $F_2$ using $Q^2$ or $\tilde W^2$ as the choice of 
scale. The curious `dip' of the MRST curve is due to the suppressed
gluon contribution at low $x$ while the smoother behaviour of the 
fit with the $z$-dependent scale reflects a dominance of the gluon
contribution at small $x$.
  
In summary, the effects of resumming some of the $\log(1-x)$ terms through
the change of scale does not lead to any improvement phenomenologically
though it is consistent with a significantly different, and perhaps
rather more natural, form of the gluon distribution at very low values 
of $Q^2$. The exercise indicates that there is therefore a degree of 
uncertainty in the detailed nature of the gluon density at such low scales.

\noindent{\bf Acknowledgment}

I am grateful to Robert Thorne for discussions.

\newpage  
          
\newpage

\newpage  

\newpage

\end{document}